\documentclass[apj]{emulateapj}

\shorttitle{Clustering of X-ray AGN at z$\sim$3.4}
\shortauthors{Allevato et al.}

\begin{document}

\slugcomment{Accepted for publication in The Astrophysical Journal}

\title{The \textit{Chandra} COSMOS Legacy Survey: Clustering of X-ray selected AGN at 2.9$\leq$\MakeLowercase{z}$\leq$5.5 using photometric redshift Probability Distribution Functions}

\author{V. Allevato\altaffilmark{1,2}, 
F. Civano\altaffilmark{3,4}
A. Finoguenov\altaffilmark{1,2}, 
S. Marchesi\altaffilmark{3,4,5},
F. Shankar\altaffilmark{6},
G. Zamorani \altaffilmark{7}, 
G. Hasinger\altaffilmark{8},
M. Salvato\altaffilmark{9},
T. Miyaji\altaffilmark{10},
R. Gilli \altaffilmark{5} ,
N. Cappelluti\altaffilmark{3}, 
M. Brusa\altaffilmark{5,6}, 
H. Suh\altaffilmark{8},
G. Lanzuisi\altaffilmark{5,6},  
B. Trakhtenbrot\altaffilmark{11},
R. Griffiths\altaffilmark{12},
C. Vignali\altaffilmark{5,6},
K. Schawinski\altaffilmark{11},
A. Karim\altaffilmark{13}}

\altaffiltext{1}{Department of Physics, University of Helsinki, Gustaf H\"allstr\"omin katu 2a, FI-00014 Helsinki, Finland}
\altaffiltext{2}{University of Maryland, Baltimore County, 1000 Hilltop Circle, Baltimore, MD 21250, USA}
\altaffiltext{3}{Yale Center for Astronomy and Astrophysics, 260 Whitney Avenue, New Haven, CT 06520, USA}
\altaffiltext{4}{Harvard Smithsonian Center for Astrophysics, 60 Garden Street, Cambridge, MA 02138, USA}
\altaffiltext{5}{Dipartimento di Fisica e Astronomia, Alma Mater Studiorum, Università di Bologna, viale Berti Pichat 6/2, 40127, Bologna}
\altaffiltext{6}{Department of Physics and Astronomy, University of Southampton, Highfield, SO17 1BJ, UK}
\altaffiltext{7}{INAF-Osservatorio Astronomico di Bologna, Via Ranzani 1, 40127 Bologna, Italy}
\altaffiltext{8}{Institute for Astronomy, University of Hawaii, 2680 Woodlawn Drive, Honolulu, HI 96822, USA}
\altaffiltext{9}{Max-Planck-Institute f\"ur Extraterrestrische Physik, Giessenbachstrasse 1, D-85748 Garching, Germany}
\altaffiltext{10}{Instituto de Astronomia, Universidad Nacional Autonoma de Mexico, Ensenada, Mexico (mailing adress: PO Box 439027, San Ysidro, CA, 92143-9024, USA)}

\altaffiltext{11}{Institute for Astronomy, Department of Physics, ETH Zurich, Wolfgang-Pauli-Strasse 27, CH-8093 Zurich, Switzerland}
\altaffiltext{12}{University of Hawaii at Hilo, 200 W.Kawili St., Hilo, HI 96720}
\altaffiltext{13}{Argelander-Institut f\"ur Astronomie, Universit\"at Bonn, Auf dem H\"ugel 71, D-53121 Bonn, Germany}

\begin{abstract}

We present the measurement of the projected and redshift 
space 2-point correlation function (2pcf) of the new catalog of \textit{Chandra} 
COSMOS-Legacy AGN at 2.9$\leq$z$\leq$5.5 ($\langle L_{bol} \rangle 
\sim$10$^{46}$ erg/s) using the generalized clustering 
estimator based on phot-z probability distribution functions (Pdfs) in addition to any available spec-z.
We model the projected 2pcf estimated using $\pi_{max}$ = 200 h$^{-1}$ Mpc 
with the 2-halo term and we derive a bias 
at z$\sim$3.4 equal to b = 6.6$^{+0.60}_{-0.55}$, which 
corresponds to a typical mass of the hosting halos 
of log M$_h$ = 12.83$^{+0.12}_{-0.11}$ 
h$^{-1}$ M$_{\odot}$. A similar bias
is derived using the redshift-space 2pcf, modelled including the typical phot-z error 
$\sigma_z$ = 0.052 of our sample at z$\geq$2.9.  
Once we integrate the projected 2pcf up to $\pi_{max}$ = 200 h$^{-1}$ Mpc, the bias of 
XMM and \textit{Chandra} COSMOS at z=2.8 used in Allevato et al. (2014) is consistent with our 
results at higher redshift.
The results suggest only a slight increase of the bias factor of COSMOS AGN at z$\gtrsim$3
with the typical hosting halo mass of moderate luminosity
AGN almost constant with redshift and equal to 
logM$_h$ = 12.92$^{+0.13}_{-0.18}$ at z=2.8 and log M$_h$ = 12.83$^{+0.12}_{-0.11}$ 
at z$\sim$3.4, respectively.
The observed redshift evolution of the bias of COSMOS 
AGN implies that moderate luminosity AGN still inhabit group-sized halos
at z$\gtrsim$3, but slightly less massive than
observed in different independent studies using X-ray AGN at z$\leq2$.

\end{abstract}


\section{Introduction}
\label{sec:intro}

\begin{figure*}
\plottwo{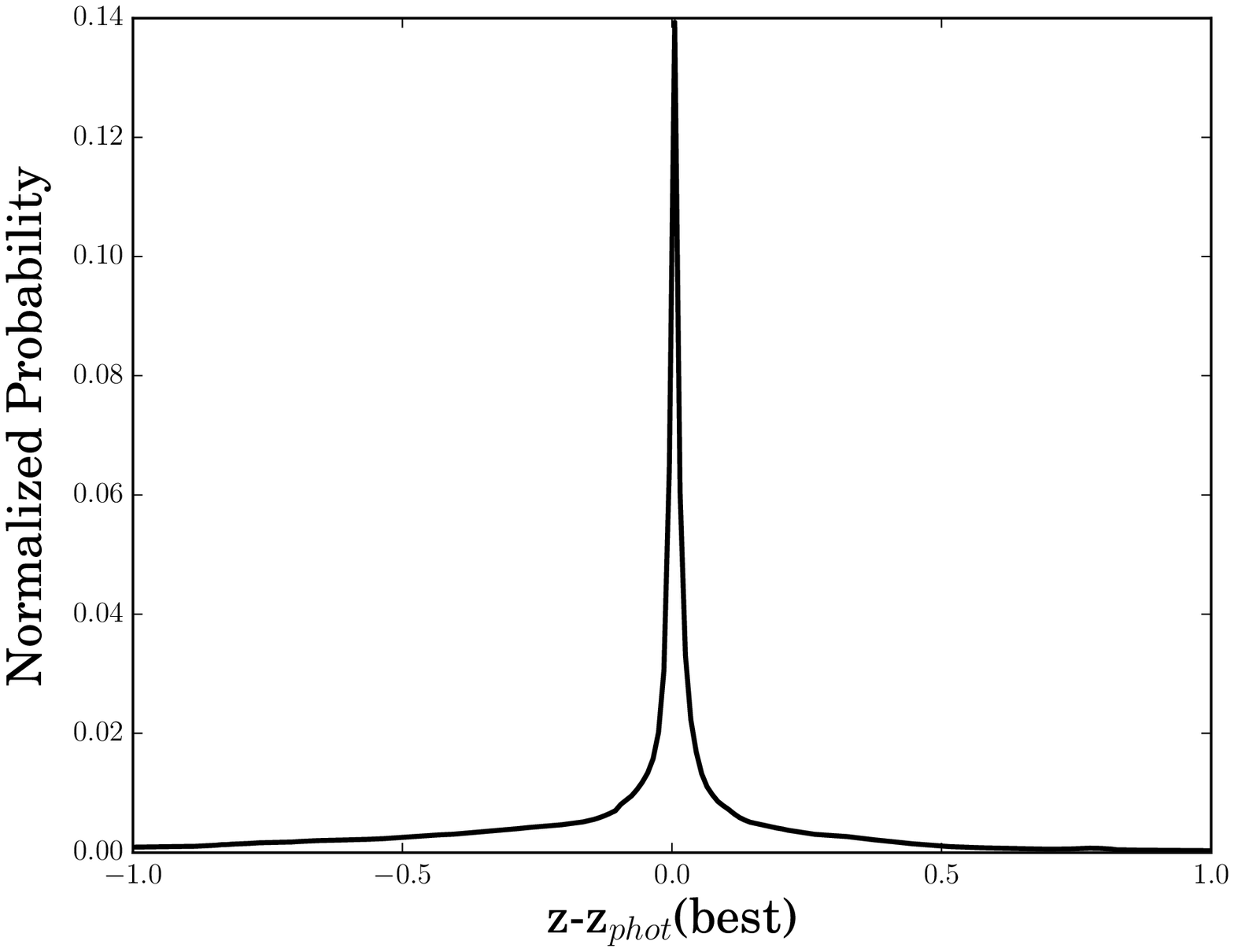}{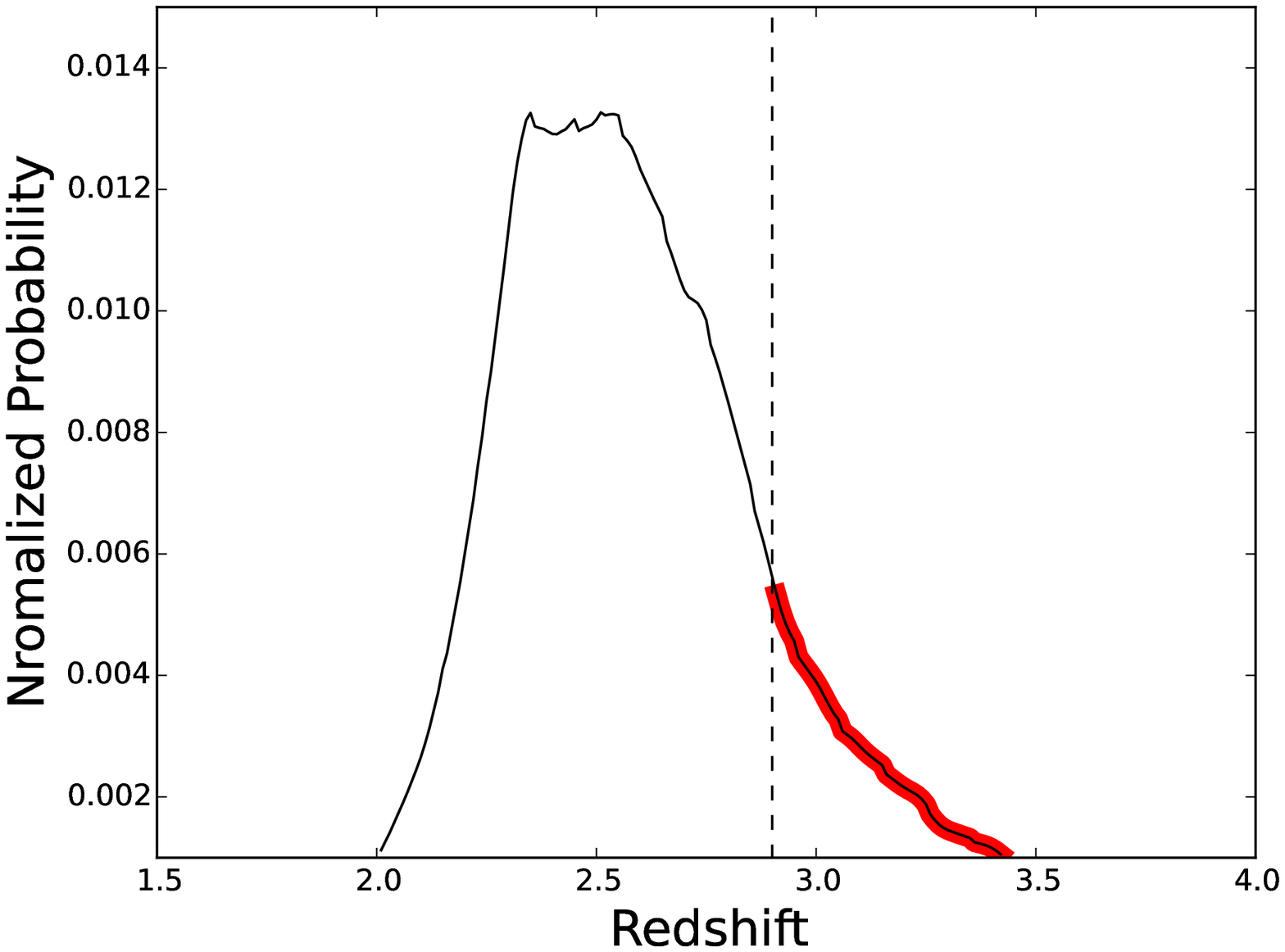}
\caption{\footnotesize Left Panel: Mean normalized phot-z Pdf of all CCL AGN with best-fit phot-z $>$ 2.9. Right Panel: Normalized phot-z Pdf for the source lid766. This source has a best-fit photo-z value $<$ 2.9, but a phot-z Pdf(z$_i >$2.9)$>$0.001 (red thick line). The redshifts above this threshold, weighted by their Pdf, have been taken in account in the catalog used to estimate the 2pcf.} 
\label{fig2}
\end{figure*}

The presence of a nuclear supermassive black hole (BH) in 
almost all galaxies in the present day
Universe is an accepted paradigm in astronomy (e.g. Kormendy \& Richstone 1995;
Kormendy \& Bender 2011).
Despite major observational and theoretical efforts over the last two
decades, a clear explanation for the origin and evolution of BHs and
their actual role in galaxy evolution remains elusive. Diverse
scenarios have been proposed. One possible picture includes major
galaxy merger as the main triggering mechanism (e.g. Hopkins et al. 2006; Volonteri et al. 2003, Menci et al.
2003,2004). On the other hand, there is
mounting observational evidence suggesting that moderate levels of
AGN activity might not be always causally connected to galaxy
interactions (Lutz et al. 2010, Mullaney et al. 2012, Rosario et al. 2013, 
Villforth et al. 2014). 
Several works on the morphology of the 
AGN host galaxies suggest that, even at moderate luminosities,
a large fraction of AGN is not associated with morphologically 
disturbed galaxies. This trend has been observed both at low 
(z $\sim$ 1, e.g., Georgakakis et al. 2009; Cisternas et al. 2011) 
and high (z $\sim$ 2, e.g., Schawinski et al. 2011, 2012; Kocevski et al. 2012, 
Treister et al. 2012) redshift.
Theoretically, in-situ processes, such as disk
instabilities or stochastic accretion of gas clouds, have also been
invoked as triggers of AGN activity (e.g. Genzel et al. 2008, 
Dekel et al. 2009, Bournaud et al. 2011).

AGN clustering analysis provides a unique way to unravel 
the knots of this complex situation, providing important, independent 
constraints on the BH/galaxy formation and co-evolution.
In the cold dark matter-dominated Universe galaxies and their BHs are believed to populate
the collapsed dark matter halos, thus reflecting the spatial distribution of dark matter in the
Universe. The most common statistical estimator for large-scale clustering is the two-point
correlation function (2pcf, Davis \& Peebles 1983). This quantity measures the excess
probability above random to find pairs of galaxies/AGN separated by a given scale $r$. By
matching the observed 2pcf to detailed outputs of dark matter numerical simulations, one
can infer the typical mass of the hosting dark matter halos. This is derived 
through the so called AGN bias b, enabling then to pin down the typical environment where AGN
live. This in turn can provide new insights into the physical mechanisms responsible for
triggering AGN activity.

The 2pcf of AGN has been measured in optical 
large area surveys, such as the 2dF (2QZ, Croom et al. 2005; 
Porciani \& Norberg 2006) and the Sloan Digital Sky Survey (SDSS, 
Li et al. 2006; Shen et al. 2009; Ross et al. 2009). 
These optical surveys are thousands of square degree 
fields, mainly sampling rare and high luminosity
quasars. The amplitude of the 2PCF of quasars 
suggests that these luminous AGN are
hosted by halos of roughly constant mass, a 
few times 10$^{12}$ M$_{\odot}$, out to z =3-4 (Shanks et al. 2011). 
Models of major mergers between gas-rich galaxies appear to naturally reproduce the 
clustering properties of optically selected quasars as a
function of luminosity and redshift (Hopkins et al. 2007a, 2008; Shen 2009; 
Shankar et al. 2010; Bonoli et al. 2009). This supports the scenario 
in which major mergers dominate the luminous quasar population 
(Scannapieco et al. 2004; Shankar et al. 2010; Neistein \& Netzer 2014; 
Treister et al. 2012). 

\textit{Chandra} surveys have contributed significantly to the study of the AGN clustering (e.g.
CDFS-N, Gilli et al. 2009; Chandra/Bootes, Starikova et al. 2011, Allevato et al. 2014). 
Deep X-ray data can be used to draw conclusions on the faint portions of the AGN
luminosity function, where a significant fraction of obscured sources is present. In particular, the \textit{Chandra}
survey in the two square degree COSMOS field (C-COSMOS, Elvis et al. 2009, Civano et al.
2012; \textit{Chandra} COSMOS Legacy Survey, Civano et al. 2016) has allowed the investigation of the redshift
evolution of the clustering properties of X-ray AGN, for the first time up to z$\sim$3. 
Interestingly, over a broad redshift range
(z $\sim$ 0 - 2) moderate luminosity AGN occupy DM halo masses
of log M$_h$ $\sim$ 12.5-13.5 M$_{\odot}$ h$^{-1}$. 
The clustering strength of X-ray
selected AGN has been measured by independent studies to be 
higher than that of optical quasars. Merger models usually 
fail in reproducing the data from X-ray surveys, opening the
possibility of additional AGN triggering mechanisms 
(e.g., Allevato et al. 2011; Mountrichas \& Georgakakis 2012) and/or 
multiple modes of BH accretion (e.g., Fanidakis et al. 2013). 
Recently, Mendez et al. (2015) and Gatti et al. (2016) have suggested that selection cuts 
in terms of AGN luminosity, host galaxy properties and redshift interval, might 
have a more relevant role in driving the 
differences often observed in the bias factor inferred from different surveys. 
%and therefore
%cluster like groups of galaxies.
%The general picture at z$<$2 is that the bias of moderate 
%luminosity X-ray AGN increases with redshift 
%tracing a constant group-sized halo mass. 
\begin{figure*}
\plottwo{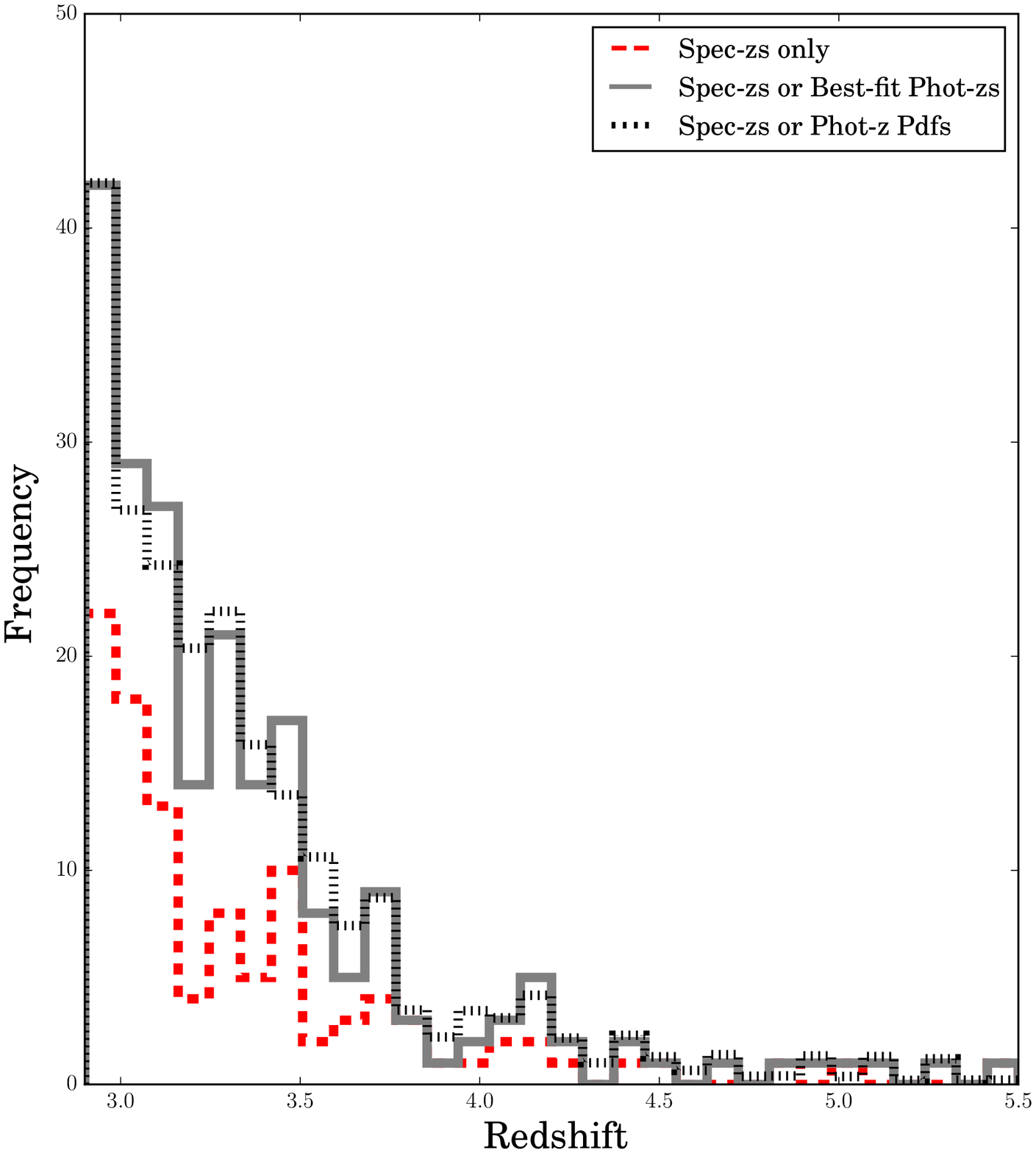}{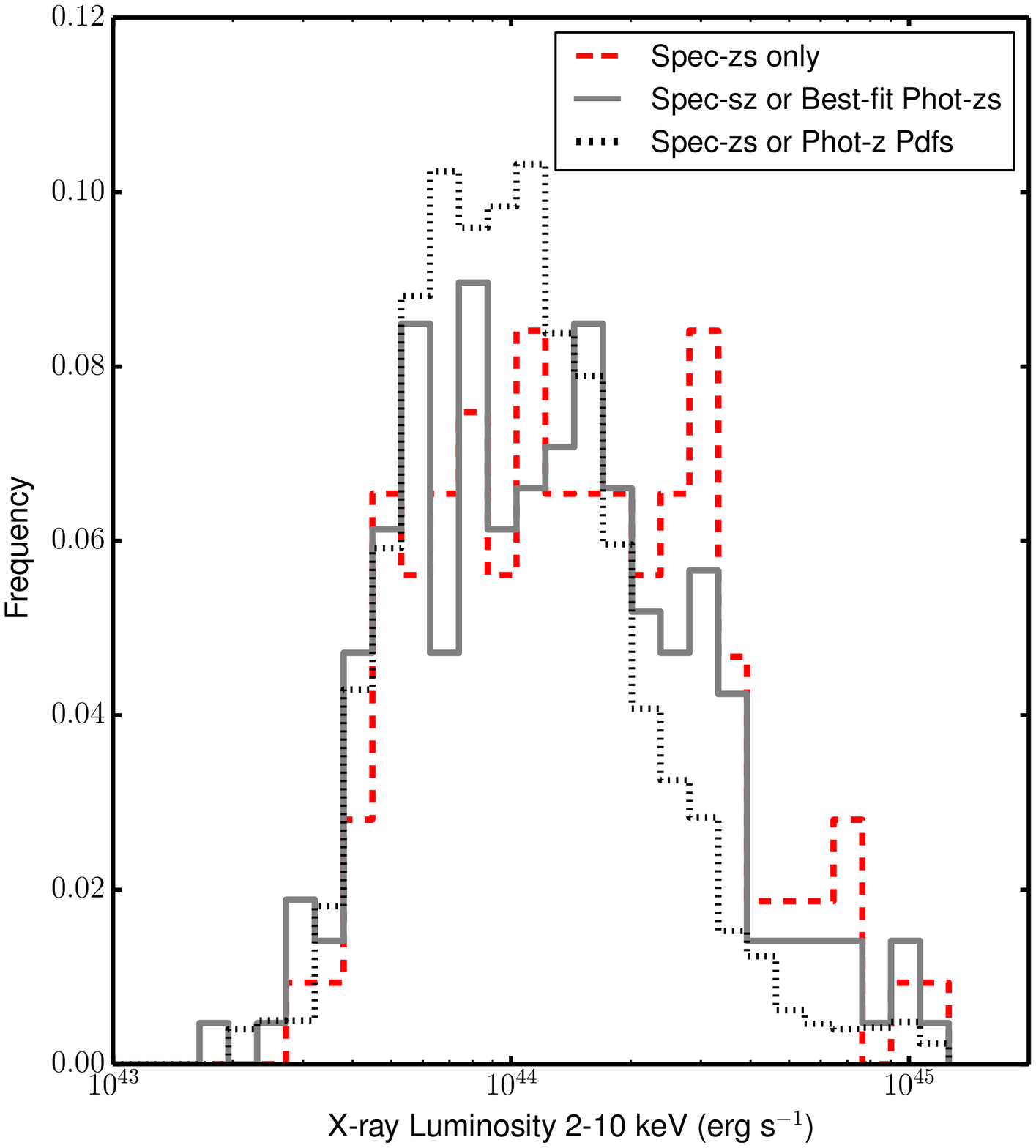}
\caption{\footnotesize Redshift (Left Panel) and 2-10 keV X-ray luminosity (Right Panel) distribution for 107 CCL AGN with known spec-z (red dashed line), 212 AGN with known spec or best phot-z (solid grey line) and 221.6 AGN with known spec-z or phot-z weighted by the Pdf (black dotted line), at 2.9$\leq$z$\leq$5.5} 
\label{fig1}
\end{figure*}
The measurement of the AGN bias is crucial at high redshifts,
especially at z$>$2-3, i.e. at the peak in the accretion history of the Universe.
At z$>$3, Shen et al. (2007,2009) measured for the first time the 2pcf of luminous 
SDSS-DR5 quasars (log L$_{bol} \sim 10^{47}$ erg/s) at $\langle z \rangle$ = 3.2 and 3.8.
Even if with very large uncertainty, they found that these objects live in 
massive halos of the order of 10$^{13}$ M$_{\odot}$ h$^{-1}$. This result is consistent with
models invoking galaxy major mergers as the main triggering mechanism for
very luminous AGN. Recently, Eftekharzadeh et al. (2015) studying a 
sample of spectroscopically confirmed SDSS-III/BOSS quasars at 2.2$\leq$z$\leq$3.4,
performed a more precise estimation of the quasar bias at high redshift. 
They found no evolution of the bias in three redshift 
bins, with halo masses equal to 3$\times$ and $\sim$0.6$\times$ 
10$^{12}$ M$_{\odot}$ h$^{-1}$ at z$\sim$2.3 and $\sim 3$, respectively.

There are only a few attempts of measuring the clustering properties 
of X-ray AGN at z $\sim$ 3. Francke et al. (2008) estimated the bias
of a small sample of X-ray AGNs (L$_{bol}$ $\sim$ 10$^{44.8}$ erg s$^{-1}$)
in the Extended Chandra Deep Field South (ECDFS), 
with very large uncertainty. They found indications that 
X-ray ECDFS AGNs reside in dark matter halos with
minimum mass of log M$_{min}$ = 12.6$^{+0.5}_{-0.8}$ h$^{-1}$ M$_{\odot}$.
On the other hand, Allevato et al. (2014) used a sample of 
\textit{Chandra} and XMM-Newton AGN in COSMOS 
with moderate luminosity (log L$_{bol} \sim 10^{45.3}$ erg/s) 
at $\langle z \rangle$=2.86. 
For the first time they estimated the bias of X-ray selected AGN at high redshift, 
suggesting that they inhabit halos of logM$_h$ = 12.37$\pm$0.10 M$_{\odot}$ h$^{-1}$. 
They also extended to z$\sim$3 the result that Type 1 AGN reside 
in more massive halos than Type 2 AGN.
Recently, Ikeda et al. (2015) estimated the clustering properties of low-luminosity 
quasars in COSMOS at 3.1$\leq$z$\leq$4.5, using the cross-correlation between 
Lyman-Break Galaxies (LBGs) and 25 quasars with 
spectroscopic and photometric redshifts. They derived a 86\% 
upper limit of 5.63 for the bias at z$\sim$ 4.

In this paper we want to extend the study of the clustering properties of 
X-ray selected AGN to z$>$3 using the new \textit{Chandra} COSMOS-Legacy data.
To this goal, we perform clustering measurements using techniques based on photometric redshift 
in the form of probability distribution functions (Pdfs), in addition to any available
spectroscopy. This is motivated by the development in the last years of 
clustering measurement techniques based on photometric redshift Pdfs 
by Myers, White \& Ball (2009), Hickox et al. (2011, 2012) and Mountrichas et al. (2013)
and Georgakakis et al. (2014).
One of the advantages of this new clustering estimator is that one can use in
the analysis all sources not just the optically brighter ones
for which spectroscopy is available. For this reason it is well suited to clustering investigations using future large X-ray AGN surveys, where the fraction of spectroscopic redshifts might be small. 

Throughout the paper, all distances are 
measured in comoving coordinates and are given in
units of Mpc $h^{-1}$, where $h=H_0/100$km/s. 
We use a $\Lambda$CDM cosmology with
$\Omega_M=0.3$, $\Omega_\Lambda=0.7$, 
$\Omega_b=0.045$, $\sigma_8=0.8$.
The symbol $log$ signifies a base-10 logarithm.

\begin{deluxetable*}{lllllllll}
\tabletypesize{\scriptsize}
\tablewidth{0pt}
\tablecaption{Properties of the AGN Samples \label{tbl-1}}
\tablehead{
\colhead{Sample} &
\colhead{ N } & 
\colhead{ $\Sigma$Pdf$_j$(z$\geq$2.9)} & 
\colhead{$\langle z \rangle$} &
\colhead{$log \langle L_{bol} \rangle$}  &
\colhead{$b$} &
\colhead{logM$_h$} &
\colhead{$b$} &
\colhead{logM$_h$}\\
\colhead{} &
\colhead{} &
\colhead{} &
\colhead{} &
\colhead{erg s$^{-1}$} &
\colhead{Eq. 7} &
\colhead{h$^{-1}$M$_{\odot}$} &
\colhead{Eq. 14} &
\colhead{h$^{-1}$M$_{\odot}$} }
\startdata
Spec-zs + Phot-z Pdfs & 457 & 221.6  & 3.36\tablenotemark{a}  & 45.99$\pm$0.53 & 6.6$^{+0.6}_{-0.55}$ & 12.83$^{+0.12}_{-0.11}$ & 6.53$^{+0.52}_{-0.55}$ & 12.82$^{+0.11}_{-0.13}$\\
%(5.56$\pm$0.7)\tablenotemark{*} & 
Spec-zs + Best-fit Phot-zs & 212 & 212 & 3.34 & 45.93$\pm$0.17 & 6.48$^{+1.27}_{-1.36}$ & 12.81$^{+0.24}_{-0.35}$ & 6.96$^{+0.72}_{-0.73}$ & 12.90$^{+0.15}_{-0.15}$\\
Spec-zs only & 107 & 107 & 3.35 & 45.92$\pm$0.34 & 7.5$^{+1.6}_{-1.7}$ & 13.0$^{+0.25}_{-0.35}$ & 7.98$^{+1.4}_{-1.5}$ & 13.08$^{+0.22}_{-0.25}$\\

%(3.21$\pm$0.72)\tablenotemark{*} & 
\enddata
\tablenotetext{a}{Mean redshift of the sample weighted by the Pdfs.}
%\tablenotetext{*}{Excluding in the fit the negative data point(s).}

\end{deluxetable*}

\section{AGN sample at 2.9$\leq$\MakeLowercase{z}$\leq$5.5}\label{sec:AGNcat}

The \textit{Chandra}-COSMOS-Legacy survey (CCLS) is
the combination of the 1.8 Ms C-COSMOS survey (Elvis et al
2009) with 2.8 Ms of new Chandra ACIS-I observations
(Civano et al. 2016) for a total coverage of 2.2 deg$^2$ of the 
COSMOS field (Scoville et al. 2007). The X-ray source catalog consists 
of 4016 sources. 2076 ($\sim$52\%) have a secure spectroscopic 
redshift (spec-z) and for $\sim$96\% the photometric redshift (photo-z) is available.
As shown in Marchesi et al. (2016a),
the spectroscopic redshifts have been obtained with
different observing programs,
as the zCOSMOS survey (Very Large Telescope/VIMOS;
Lilly et al. 2007) and the Magellan/IMACS survey
(Trump et al. 2007, 2009). Other programs, many of which
have been specifically targeting the CCLS have been carried with Keck-MOSFIRE (P.I. F.
Civano, N. Scoville), Keck-DEIMOS (P.I.s Capak, Kartaltepe, 
Salvato, Sanders, Scoville, Hasinger), Subaru-
FMOS (P.I. J. Silverman), VLT-FORS2 (P.I. J. Coparat)
and Magellan-PRIMUS (public data).
%\textbf{21/107 ($\sim20$\%) AGN with known spec-zs have quality flag = 1.5 
%which implies lower quality spectra but with known phot-z such that 
%$\sigma_{\Delta z/(1+z_{spec})}<$0.1.}

The photo-zs are estimated following
the procedure described in Salvato et al. (2011).
Following Marchesi et al. (2016b), the accuracy of 
the photometric redshifts with respect to the whole spectroscopic redshift sample is
$\sigma_{\Delta z/(1+z_{spec})}$=0.02, with a fraction of outliers $\simeq$11\%.
At z$>$2.9 there are 9 outliers ($\Delta$z/(1 + z$_{spec}$) $>$ 0.15),
but for the remaining sources the agreement between
spec-z and photo-z has the same quality of the whole
sample. In detail, the normalized median absolute deviation $\sigma_{NMAD}$=1.48$\times$median($\parallel zspec-
zphot \parallel /(1+zspec))=0.012$.
 
%We restrict the analysis to 212 AGN detected in the full band,
%with spec or best-fit phot-zs > 2.9.
The CCL AGN sample at 2.9$<$z$<$5.5 consists of 212 AGN detected in the 0.5-10 keV band, 107/212 with spec-zs and 105/212 with only phot-zs.
To each of the 105 AGN with best-fit phot-z in the range 2.9$\leq$z$<$5.5, 
is associated a probability distribution function (Pdf), which 
gives the probability of the source to be in the redshift range 
z$_i \pm \Delta z/2$ with a binsize of $\Delta$z = 0.01. 
The integrated area of the Pdf on all redshift bins z$_i$ is normalized to 1,
i.e. $\sum_i Pdf(z_i) = 1$ for each AGN. We take into account for 
this analysis the redshift bins z$_i$ with Pdf(z$_i$) $>$0.001. Figure 1 shows 
the mean normalized phot-z Pdf for 105 sources with best-fit phot-z 2.9$\leq$z$<$5.5. 
The effective contribution to the number of AGN at z$\geq$2.9 of these 105 AGN weighted by the Pdf is 78.32 sources, i.e. 
$\sum_{j=1}^{105} Pdf_{j}(z\geq2.9)$=78.32. 

In the CCL sample there are also 246
sources with phot-z $<$ 2.9 but which
contribute to the Pdf at 2.9$\leq$ z $\leq$ 5.5 (i.e., with Pdf$>$0.001 at
2.9 $\leq$ z$_{i}$ $\leq$5.5 (see for example the Pdf of source lid766, whose nominal best-fit phot-z value is 2.51, in Figure 1). 
All these 246 sources have been taken in account in our analysis, using for each of them the Pdf of each bin of redshift 2.9 $\leq$ z$_{i}$ $\leq$5.5. 
The weighted contribution of these sources, i.e. the sum of all
weights, is equal to 36.3 AGN ($\sum_{j=1}^{246}Pdf_{j}(z\geq2.9)$=36.3). 
To all the 107 sources with known spec-z we assign a Pdf$_j$ = 1 to 
the spec-z value ($\sum_{j=1}^{107}Pdf_j$=107).
To summarize, the total effective number of CCL AGN at 2.9$\leq$z$<$5.5 weighted by the Pdf and used 
for the clustering measurements is 78.3 + 36.3 + 107 = 221.6 objects.

Figure 2 shows the normalized redshift and 2-10 keV rest-frame X-ray luminosity distribution for our  
sample of CCL AGN at 2.9$<$z$<$5.5, when 
the phot-z Pdfs are used (black dotted line, $\langle z \rangle$=3.36). The mean 
bolometric luminosity of this sample derived using the bolometric correction defined in Equation (21) of Marconi et al. (2004) is 
$log \langle L_{bol}\rangle$ = 45.99 erg s$^{-1}$.
For comparison, we also show the normalized distributions of our 
AGN sample when only the best-fit phot-zs are taken into account in addition 
to any available spec-z (gray solid line, $\langle z \rangle$=3.34) and 
for the sample with known spec-z (red-dashed line, $\langle z \rangle$=3.35).

\section{2PCF using phot-\MakeLowercase{z} PDFs}

\subsection{Projected 2pcf}

The most commonly used quantitative measure of large scale structure
is the 2pcf, $\xi$(r), which traces the amplitude of AGN clustering as a function of scale. $\xi$(r) is defined as a measure of the excess probability $dP$, above what is expected for an unclustered random Poisson distribution, of finding an AGN in a volume element $dV$ at a separation $r$ from
another AGN:
\begin{equation}
dP = n[1 + \xi(r)]dV
\end{equation}
where $n$ is the mean number density of the AGN sample 
(Peebles 1980). Measurements of $\xi$(r) are generally performed in comoving space, with
$r$ having units of h$^{-1}$ Mpc.

\begin{figure*}
\plottwo{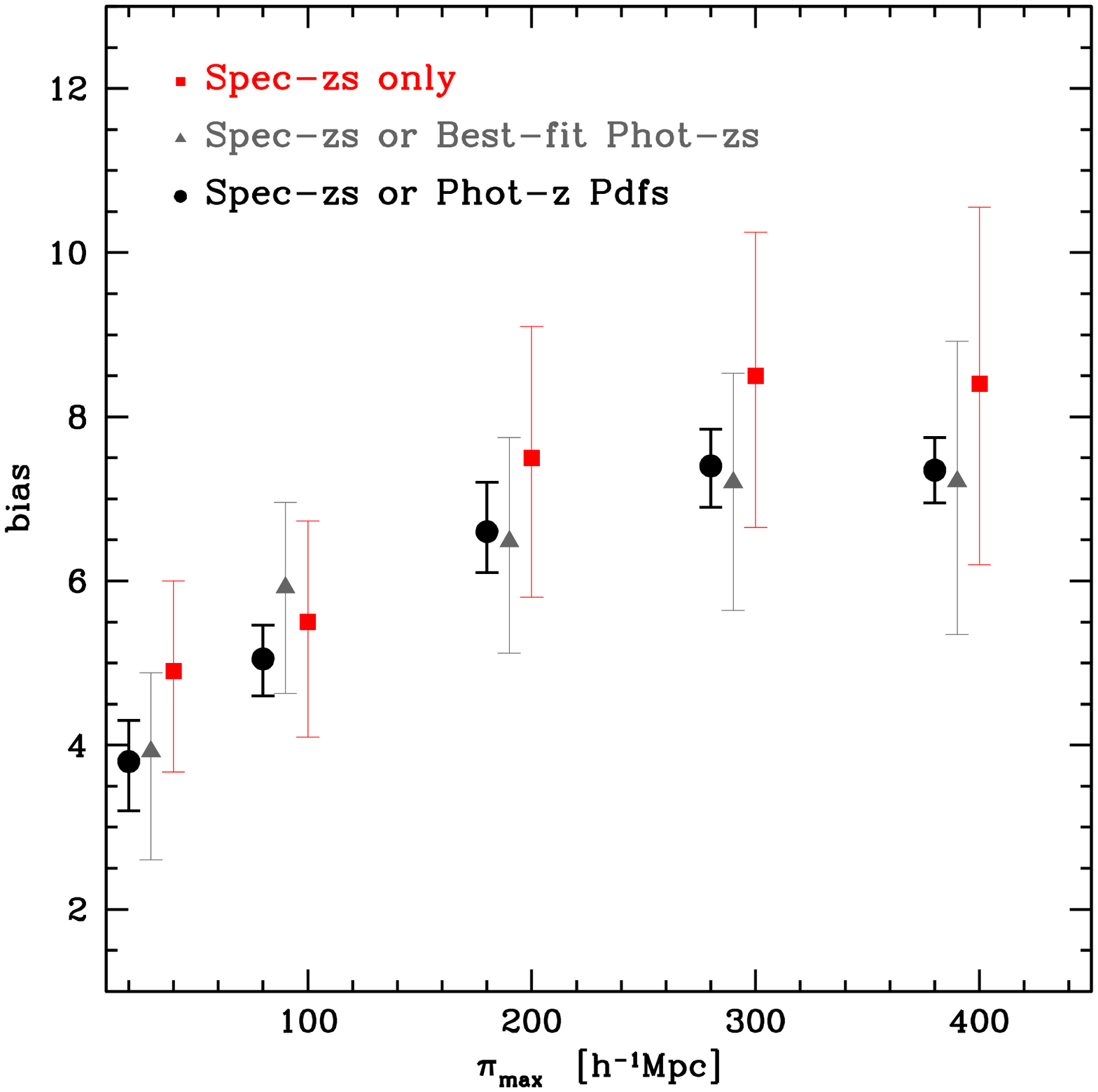}{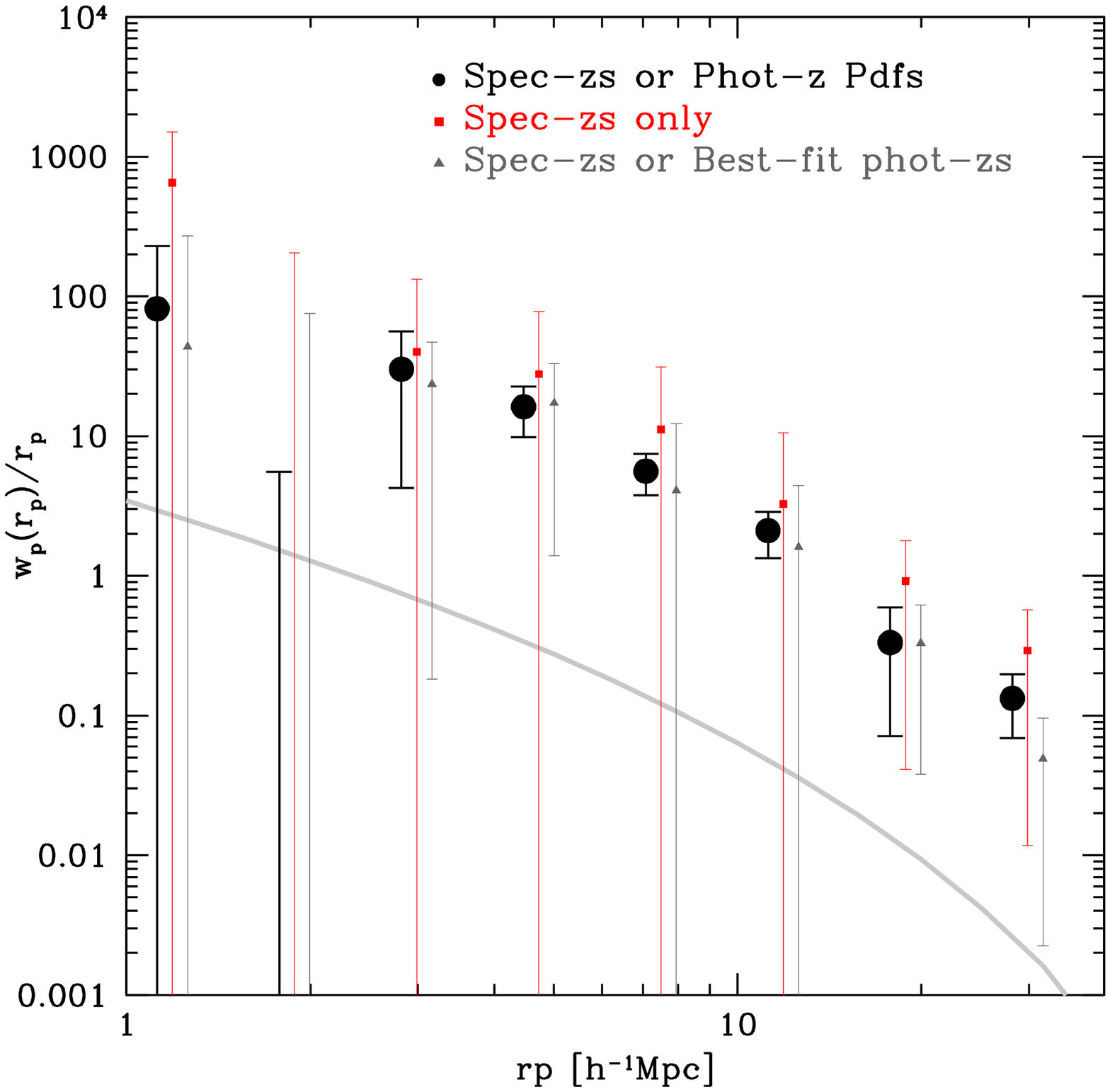}
\caption{\footnotesize  Left Panel: Bias as a function of $\pi_{max}$ for 221.6 CCL AGN at 2.9$\leq$z$<$5.5, when the projected 2pcf is measured using the generalized clustering estimator based on phot-z Pdfs in addition to any available spec-z (black dots). For comparison, the gray triangles and red squares show the bias of 212 AGN with known spec or best-fit phot-z and 107 AGN with known spec-zs at 2.9$<$z$<$5.5 when the classic LS estimator (i.e. no phot-z Pdfs) is used. Right Panel: Projected 2pcf for $\pi_{max}$=200 h$^{-1}$ Mpc. The 1$\sigma$ errors on w$_p$(r$_p$) are the square root of the diagonal components of the covariance matrix. The continuous line represents the DM projected 2pcf estimated at $\langle z \rangle$=3.36.} 
\label{fig1}
\end{figure*}

With a redshift survey, we cannot directly
measure $\xi(r)$ in physical space, because peculiar motions of galaxies
distort the line-of-sight distances inferred from redshift. To separate the
effects of redshift distortions, the spatial correlation function is
measured in two dimensions $r_p$ and $\pi$, where $r_p$ and $\pi$ are the
projected comoving separations between the considered objects in the
directions perpendicular and parallel, respectively, to the mean
line-of-sight between the two sources.  Following \citet{Dav83},
$r_1$ and $r_2$ are the redshift positions of a pair of objects, $s$ is the
redshift-space separation $(r_1 - r_2)$, and $l = \frac{1}{2}(r_1+r_2)$ is
the mean distance to the pair. The separations between the two considered
objects across $r_p$ and $\pi$ are defined as:
\begin{eqnarray}
\pi & = & \frac{\textbf{s}\cdot \textbf{l}} {|\textbf{l}|}\\
r_p & = & \sqrt{(\textbf{s} \cdot \textbf{s} - \pi^2)}
\end{eqnarray}
 Redshift space distortions only affect the correlation function along the line of sight, 
 so we estimate the so-called projected correlation function $w_p(r_p)$ \citep{Dav83}:
\begin{eqnarray}\label{eq:integral}
w(\sigma) = 2 \int_0^{\pi_{max}} \xi(\sigma,\pi) d\pi 
 \end{eqnarray}
 where $ \xi(r_p,\pi)$ is the two-point correlation function in terms of
 $r_p$ and $\pi$, measured using the \citet[LS]{Lan93} estimator:
\begin{equation}\label{eq:LZ}
\xi = \frac{1}{RR'} [DD'-2DR'+RR']
\end{equation}
where DD', DR' and RR' are the normalized data-data, data-random and random-random pairs.
%\begin{eqnarray}
%DD' = \frac{DD(r_p,\pi)}{n_d(n_d - 1)}\\
%DR' = \frac{DR(r_p,\pi)}{n_dn_r}\\
%RR' = \frac{RR(r_p,\pi)}{n_r(n_r-1)}
%\end{eqnarray}
%where $DD$, $DR$ and $RR$ are the number of data-data, data-random and
%random-random pairs at separation $r_p \pm \Delta r_p$ and $\pi \pm \Delta
%\pi$ and $n_d$, $n_r$ are the total number of sources in the data and random
%sample, respectively. 

%In this classic approach of estimating the redshift-space correlation
%function, in presence of accurate spec-zs,
%when a data-data pair is found with separation $r_p,\pi$, the number of pairs with these separation, is incremented by one, i.e. DD($r_p$,$\pi$) = DD($r_p$,$\pi$) + 1.
%Following Georgakakis et al. (2014), in the generalized clustering estimator
%the number of data-data pairs is, instead, incremented by the product Pdf$_{j=1}(z_i)$ $\times$ Pdf$_{j=2}(z_i)$:
%\begin{equation}
%DD(\sigma,\pi) = DD(\sigma,\pi) + Pdf_{k1}(z=z_i) Pdf_{k2}(z=z_j)
%\end{equation}\label{eq:prob}
%where Pdf$_{j=1}(z_i)$ is the value of the Pdf in the redshift bin 
%$z_i$ for source 1 and Pdf$_{j=2}(z_i)$ is the value of the 
%Pdf in the redshift bin $z_i$ for source 2. 

In this classic approach of estimating the redshift-space correlation
function, in presence of accurate spec-zs,
when a data-data pair with separation ($r_p,\pi$) is found, 
the pair number is incremented by one, i.e. DD($r_p$,$\pi)$ = DD($r_p,\pi)$ + 1.
Following Georgakakis et al. (2014), in the generalized clustering estimator
the number of data-data pairs with projected and line of sight 
separation ($r_p,\pi$) is, instead, incremented by the product Pdf$_{1}(z_i)$ $\times$ Pdf$_{2}(z_j)$:
\begin{equation}
DD(r_p,\pi) = DD(r_p,\pi) + Pdf_{1}(z_i) \times Pdf_{2}(z_j)
\end{equation}\label{eq:prob}
where Pdf$_{1}(z_i)$ and Pdf$_{2}(z_j)$ are the Pdf values (per redshift bin)
of the source 1 at $z=z_i$ and of the source 2 at $z=z_j$ respectively.

The measurements of the 2pcf requires the construction of a random 
catalog with the same selection criteria and observational effects as the data.
To this end, we constructed a random catalog 
where each simulated source is placed at a random position in the sky, with
its flux randomly extracted from the catalog of real source fluxes.
The simulated source is kept in the random sample 
if its flux is above the sensitivity map value at
that random position \citep{Miy07, Cap09}.
The corresponding redshift for each random object is then assigned based on the
%smoothed redshift distribution of the AGN data sample. 
%We assumed a Gaussian smoothing length $\sigma_z = 0.2$. This is
%a good compromise between scales 
%that are too small, which would suffer from 
%local density variations, and those that are too 
%large, which would oversmooth the distribution. 
smoothed redshift distribution of the AGN sample, 
where each redshift is weighted by the Pdf associated to that redshift for the 
particular source. Since the phot-z Pdfs are already taken into account in the generation of 
the random redshifts, we decided to assign Pdf=1 to each random source.

In the halo model approach, the large scale amplitude
signal is due to the correlation between objects in distinct halos
and the bias parameter defines the 
relation between the large scale clustering amplitude 
of the AGN correlation function and the DM 2-halo term:
\begin{equation}\label{eq:b}
b^{2-h}(r_p)=(w_{AGN}(r_p)/w_{DM}^{2-h}(r_p))^{1/2}
\end{equation}
We first estimated the DM 2-halo term at the median redshift of the sample, using:
\textbf{\begin{equation}
w_{DM}^{2-h}(r_p)=2 \int_{r_p}^{\infty} \frac{\xi^{2-h}_{DM}(r)rdr}{\sqrt{r^2-r_p^2}}
\end{equation}}
integrating up to 200 h$^{-1}$ Mpc along $\pi$, where:
\begin{equation}\label{eq:2-halo}
\xi^{2-h}_{DM}(r)=\frac{1}{2\pi^2}\int P^{2-h}(k)k^2 \left[ \frac{sin(kr)}{kr} \right]  dk
\end{equation}
$P^{2-h}(k)$ is the linear power spectrum,
assuming a power spectrum shape parameter $\Gamma = \Omega_m h =0.2$ (Efstathiou et al. 1992) 
which corresponds to $h=0.7$.

\subsection{z-space correlation function}

%\textbf{Similarly, we measured the z-space correlation function $\xi(s)$ in the range s = 1-50 h$^{-1}$Mpc, for 221.6 Chandra COSMOS-Legacy AGN at z$sim$3.4 using the phot-z Pdfs in addition to any available specz and for 107 and 107+105 AGN using the classic clustering estimator (see Figure ?).}

Similarly, we can estimate the z-space correlation function $\xi(s)$ using Equation (5) and (6), written now as a function of the redshift-space separation $s=(r_1-r_2)$ between the sources. $\xi(s)$ is affected by perturbations in the cosmological redshifts due to peculiar velocities and redshift errors.
The z-space power spectrum can be modelled in polar coordinates as follow (e.g. Kaiser 1987, Peacock et al. 2001):
\begin{equation}
P(k,\mu) = P_{DM}(k)(b+f\mu^2)^2exp(-k^2\mu^2\sigma^2)
\end{equation}
where $k=\sqrt{k^2_{\bot}+k^2_{\|}}$, $k_{\bot}$ and $k_{\|}$ 
are the wavevector components perpendicular and parallel to the line of sight, respectively. $\mu$ = $k_{\bot}$/$k_{\|}$, P$_{DM}(k)$ is the dark matter power spectrum, $b$ is the linear bias factor, $f$ is the growth rate of density fluctuations. $\sigma$ is the displacement along the line of sight due to random perturbations of cosmological redshifts.Assuming standard gravity, we approximated the growth rate $f \simeq \Omega_{M}(z)^{\gamma}$, with $\gamma$ = 0.545 (e.g. Sereno et al. 2015). 

The $f\mu^2$ term parametrizes the coherent motions due to large-scale structures, enhancing the clustering signal on all scales. The exponential cut-off term describes the random perturbations of the redshifts caused by both non-linear stochastic motions and redshift errors.
The integration of Eq.(7) over the angle $\mu$, and then the Fourier anti-transformation gives:
\begin{equation}
\xi(s) = b^2 \xi^{'}(s) + b\xi^{''}(s) + \xi^{'''}(s)
\end{equation}
The main term, $\xi^{'}(s)$, is the Fourier anti-transform of the monopole $P^{'}(k)$:
\begin{equation}
P^{'}(k) = P_{DM}(k) \frac{\sqrt{\pi}}{2k \sigma} erf(k \sigma),
\end{equation}\label{eq:pkmono}
that corresponds to the model given by Equation (10) when neglecting the dynamic distortion term.

In our case, photo-z errors perturb the most the distance measurements along the line of sight. Therefore the small-scale random motions are negligible with respect to photo-z errors.
The cut-off scale in Eq. (12) can thus be written as:
\begin{equation}
\sigma = \frac{c \sigma_z}{H(z)}
\end{equation}
where H(z) is the Hubble function computed at the median redshift
of the sample, and $\sigma_z$ is the typical photo-z error.

In this case, knowing the cut-off scale, the AGN bias can be derived from the Fourier anti-transform 
of the monopole $P^{'}(k)$, i.e.:
\begin{equation}
b^2 = \frac{\xi^{'}(s)}{\xi(s)}
\end{equation}
where $\xi(s)$ is the observed z-space 2pcf of our AGN sample.

\section{Results}

\subsection{w$_p(r_p)$ and Bias}

We have measured the 2pcf of 221.6 CCL AGN at 
2.9$\leq$z$<$5.5, using the generalized clustering estimator defined in Equation (6),
based on phot-z Pdfs in addition to any available spec-z. The projected 
2pcf w$_p(r_p)$ is then estimated using Equation (4).
%For 107 AGN the spec-z is available and the Pdf is set to unity
%($\Sigma_{j=1}^{107}$Pdf$_j$=107). For 105 AGN 
%with best-fit phot-z in the range 2.9-5.5, $\Sigma_{j=1}^{105}$Pdf$_j$=78.32 while for
%246 with best-fit phot-z$<$2.9 but with 
%Pdf(z$_{i}>2.9)>$0.001, $\Sigma_{j=1}^{246}$Pdf$_j$=36.3.

The typical value of $\pi_{max}$ used in clustering measurements 
of both optically-selected luminous quasars and X-ray selected AGN is $\sim$20-100 h$^{-1}$Mpc
(e.g. Zehavi et al. 2005, Coil et al. 2009, Krumpe et al. 2010, Allevato et al. 2011).
The optimum $\pi_{max}$ value can be determined by measuring the 2pcf 
for different $\pi_{max}$ and then adopting the value at which
the amplitude of the signal appears to level off. 

Figure 3 (Left Panel) shows the bias factor estimated 
for different values of $\pi_{max}$ in Equation (4), when the phot-z Pdfs are used in addition to any available spec-z. For comparison, we also estimated the bias for case $i$) 107 AGN with known spec-zs; case $ii$) 107+105 AGN with 
known spec-z or best-fit phot-zs. In these cases the 
2pcf is measured using the classic LS estimator and the Pdf is set to unity for 
each source. 

As expected, when including phot-zs in the analysis, the bias levels-off only at large scales, because of the large uncertainties in the redshifts measured via photometric methods (Georgakakis et al. 2014). Surprisingly, 
even if the error bars are large, an increase of the bias factor at $\pi_{max}>$100 h$^{-1}$ Mpc 
is suggested also when only spec-zs are used with the classic 2pcf estimator. This suggests that a fraction of spec-zs might be affected by large errors (see Section 4.2).
% and/or peculiar velocities are very large at z$>$3 (see Section 4.2).}

The amplitude of the projected 2pcf of 
CCL AGN measured using the generalized 
clustering estimator converges at $\pi_{max}\geq$ 200 h$^{-1}$Mpc. 
We decide to use $\pi_{max}$= 200 h$^{-1}$Mpc in order to balance the 
advantage of integrating out redshift-space distortions against 
the disadvantage of introducing noise from uncorrelated line-of-sight structure.
%These value is larger than what is typically adopted. Georgakakis et al. (2014) suggested 
%that the difference is because of the larger uncertainties of the redshifts measured
%via photometric methods. However we find a large $\pi_{max}$ value for the level-off of the 
%signal also when using a small sample of AGN with known spec-zs (see Fig. 3).}

Figure 3 (Right Panel) shows the projected 2pcf estimated using the generalized clustering estimator, 
whith $\pi_{max}$=200 h$^{-1}$Mpc. The 1$\sigma$ errors on $w_p(r_p)$ are the square root of the 
diagonal components of the covariance matrix (Miyaji et al. 2007, Krumpe et al. 2010) estimated using the bootstrap method.
The latter quantifies the level of correlation between different bins.
For comparison, we also estimate the projected 2pcf for case $i$) 107 AGN with known spec-zs; case $ii$) 107+105 AGN with 
known spec-z or best-fit phot-zs. Note that in these cases the classic LS estimator 
is used (i.e. Pdf = 1 for each source) and $\pi_{max}$ is fixed to 200 h$^{-1}$Mpc also in these cases.
%In fact, as shown in Figure 3, for higher values of $\pi_{max}$ the 2pcf is characterised by large uncertainty.

Following Eq. \ref{eq:b}, we derive the best-fit bias by using a $\chi^2$ minimization technique with 1 free parameter in the range $r_p$ = 1 - 30 h$^{-1}$ Mpc,
where $\chi^2 = \Delta^T M^{-1}_{cov} \Delta$. In detail, 
$\Delta$ is a vector composed of $w_{p}(r_p)-w_{mod}(r_p)$ (see Equations \ref{eq:integral} and \ref{eq:b}), $\Delta^T$ is its transpose
and M$^{-1}_{cov}$ is the inverse of covariance matrix.
The latter full covariance matrix is used in the fit to take into account the
correlation between errors.

\begin{figure}
\plotone{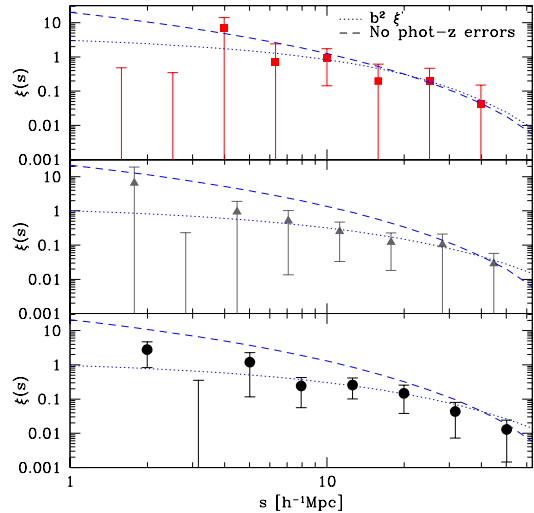}
\caption{\footnotesize Redshift-space correlation function of 221.6 CCL AGN at 2.9$<$z$<$5.5 derived using the generalized clustering estimator based on phot-z Pdfs in addition to any available spec-z (black circles in the bottom panel). For comparison, the gray triangles (Middle Panel) and the red squares (Upper Panel) show the 2pcf of 212 AGN with known spec or best-fit phot-z and 107 AGN with known spec-zs at 2.9$<$z$<$5.5 estimated using the classic LS estimator (i.e. Pdf=1 for each source). The dotted lines show the best-fit models obtained including the dominant term $b^2 \xi^{'}$ in Equation 11, while the dashed lines is the model without the photo-z damping term. The error bars show the square roots of the diagonal values of the covariance matrix.} 
\label{figxis}
\end{figure}

\begin{figure*}
\plottwo{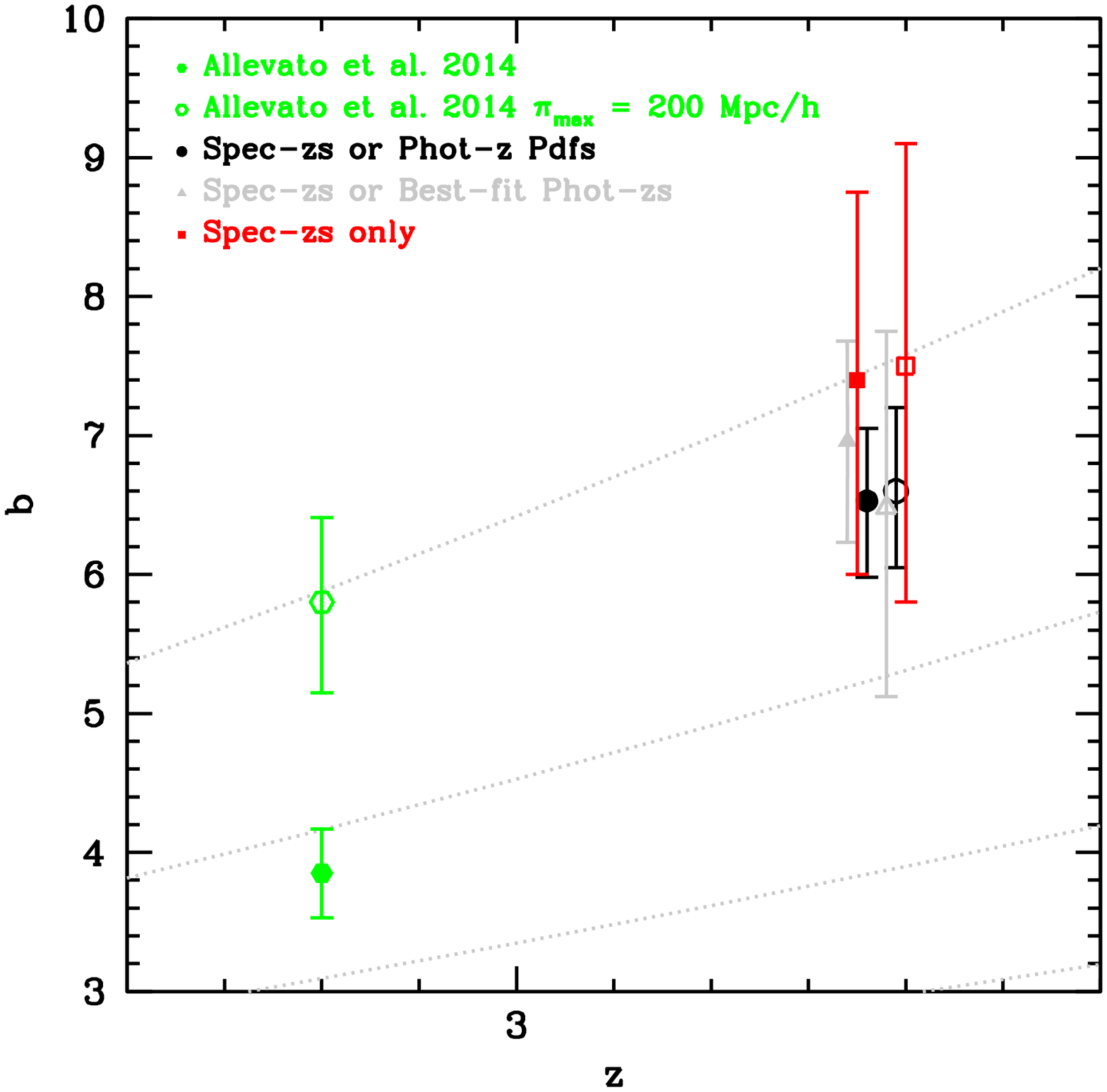}{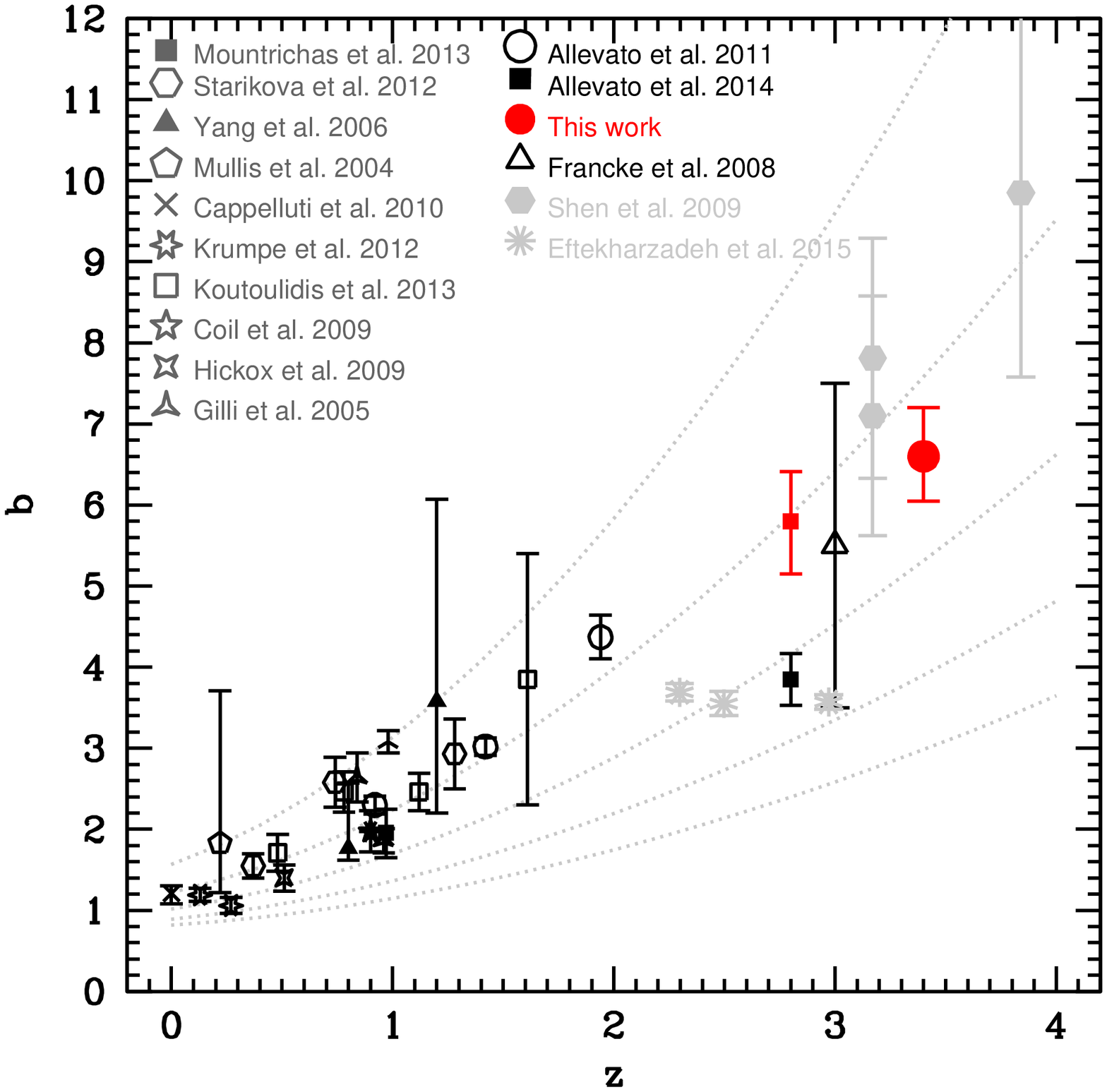}
\caption{\footnotesize Left Panel: Bias of CCL AGN at z$\sim$3.4 estimated using the projected 2pcf (empty points) and the z-space 2pcf (filled points). The empty symbols are offset in the horizontal direction by +0.04 for clarity. For comparison the bias of XMM and \textit{Chandra} COSMOS AGN at z = 2.8 is shown as presented in Allevato et al. (2014) using $\pi_{max} =$ 40 h$^{-1}$ Mpc (filled hexagon) and when the bias is re-estimated using $\pi_{max} =$ 200 h$^{-1}$ Mpc. The dashed lines show the expected b(z) of typical DM halo with mass of 12, 12.5, 13, h$^{-1}$ M$_{\odot}$ in log scale (from bottom to top), based on Sheth et al. (2001) formalism. Right Panel: Bias parameter as a function of redshift for X-ray selected AGN (black) from previous studies as described in the legend. The red circle at z$\sim$3.4 show the bias factors as estimated in this work for CCL AGN. The red square shows the bias factor as re-estimated in the present paper using the same catalog of XMM and Chandra-COSMOS AGN used in Allevato et al. (2014), but with $\pi_{max}$=200 h$^{-1}$Mpc. The dashed lines show the expected b(z) of typical DM halo with mass of 11.5, 12, 12.5, 13, 13.5 h$^{-1}$ M$_{\odot}$ in log scale (from bottom to top), based on Sheth et al. (2001). Bias factors from different studies are converted to a common cosmology ($\Omega_{\Lambda}$ = 0.7, $\Omega_m$ = 0.3, $\sigma_8$ = 0.8).} 
\label{fig5}
\end{figure*}

As shown in Table 1, 
%The best-fit bias factor is estimated in the range $r_p$ = 1 - 30 h$^{-1}$ Mpc 
we derived a bias for our sample of CCL AGN equal to 
b = 6.6$^{+0.60}_{-0.55}$ at $\langle z \rangle$=3.36.
Following the bias-mass relation $b(M_h, z)$ described 
in van den Bosch (2002) and Sheth et al. (2001), 
the AGN bias corresponds to a typical mass of 
the hosting halos of log M$_h$ = 12.83$^{+0.12}_{-0.11}$ 
h$^{-1}$ M$_{\odot}$. It is worth noticing that this is a 
typical/characteristic mass of the halos hosting CCL AGN. Only 
the HOD modelling of the clustering signal at all scales can provide 
the entire hosting halo mass distribution for this sample.

%The errors correspond to 
%$\Delta \chi^2$ = 1 using a $\chi^2$ 
%minimization technique with 1 free parameter.
The bias has a larger uncertainty when derived 
from the 2pcf estimated without the phot-z Pdfs. In detail, we find 
b = 7.5$^{+1.6}_{-1.7}$ at $\langle z \rangle$=3.35 for 107 
AGN with known spec-zs (case $i$) and 
b = 6.48$^{+1.27}_{-1.36}$ at $\langle z \rangle$=3.34 for 107+105 
AGN with known spec or best-fit phot-zs (case $ii$). Note that in these cases the 
2pcf is measured using the classic LS estimator and the Pdf is set to unity for 
each source. 
%This result suggests then an increase of the bias factor with redshift, 
%for the first time up to z$\sim$3.4 for X-ray selected AGN. 

\subsection{$\xi(s)$ and phot-z errors}

To investigate the effect of phot-z errors on the 2pcf, we also measured the z-space correlation function $\xi(s)$. Figure 4 shows $\xi(s)$ for 221.6 CCL AGN using the generalized clustering estimator based on phot-z Pdfs in addition to any available spec-z. For comparison, the gray triangles show $\xi(s)$ for 
case i) and case ii).
%212 AGN with known spec or best-fit phot-z and 107 AGN with known spec-zs estimated using the classic LS estimator (i.e. Pdf=1 for each source).

As described in section 3.2, $\xi(s)$ is affected by the Kaiser effect that enhances the clustering signal at all scales and by phot-z errors, that are modelled by using an exponential cut-off in the z-space power spectrum.
For our sample of CCL AGN the typical error on phot-zs is $\sigma_{z}$=0.012$\times(1+z_{spec})$ = 0.052 at $\langle z \rangle$=3.4. This implies a cut-off scale $\sigma$ = 43.45 h$^{-1}$Mpc (see Equation 13). 
Including the phot-z damping in the modeling of $\xi(s)$,
we derived the best-fit bias by using Equation 14 and the $\chi^2$ minimization technique with 1 free parameter in the range s=1-50 h$^{-1}$Mpc. In particular, for the z-space 2pcf measured using the generalized clustering estimator we derived b = 6.53$^{+0.52}_{-0.58}$. For the sample of 212 AGN with known spec or best-fit phot-zs for which the classic clustering estimator is used, we derived b = 6.96$^{+0.72}_{-0.73}$. These values 
are in perfect agreement with the bias obtained from the projected 2pcf with $\pi_{max}$ = 200 h$^{-1}$Mpc.
This confirms that the convergence of the projected 2pcf observed only at large scales ($\pi_{max}\geq$200 h$^{-1}$Mpc) is due to large phot-z errors. In fact, for large redshift errors and small survey area it is necessary to integrate the correlation function up to large scales to fully correct for them. 

We also estimated the bias using the z-space 2pcf for 107 AGN with known spec-zs. 
In general, we do not expect spec-zs to be affected by large errors. 
%By using the projected 2pcf we found a bias b = 6.6$^{+0.6}_{-0.5}$ for 
%$\pi_{max}$ = 200 h$^{-1}$Mpc.
If we do not include the phot-z damping in the model of $\xi(s)$, we obtain for this sample a bias 
b = 5.86$^{+1.13}_{-1.05}$, which is lower than b = 7.5$^{+1.6}_{-1.7}$ obtained 
using the projected 2pcf and $\pi_{max}$ = 200 h$^{-1}$Mpc.
A better fit of $\xi(s)$ and a larger bias can be obtained only if we 
include in the model spec-z errors of the order of
$\sigma_z$ = 0.02-0.025. In particular, for $\sigma_z$ = 0.020 (cut-off scale $\sigma$ = 16.7 h$^{-1}$Mpc) and $\sigma_z$ = 0.025 ($\sigma$ = 20.9 h$^{-1}$Mpc), we derived b = 7.4$^{+1.35}_{-1.40}$ and b = 8.01$^{+1.4}_{-1.5}$, respectively. However, given the low statistics, smaller values of $\sigma_z$ and then of the bias can not be ruled-out.
This error redshift is larger than what is expected for a spectroscopic sample of AGN. However, the presence in the spectroscopic sample of $\sim$20\% of the objects (21/107) with a low quality flag (i.e. flag = 1.5, corresponding to low quality spectra, and therefore not fully reliable redshift, but with known phot-z such that $\sigma_{\Delta z/(1+z_{spec})}<$0.1) could explain both the improvement of the fit including such an error in the analysis and the increase of the bias with increasing $\pi_{max}$ up to $\sim$200 h$^{-1}$Mpc when using only AGN with spectroscopic redshift.

\section{Discussion}
\label{sec:disc} 

In this section we compare our results with previous 
measurements using COSMOS AGN at z$\sim$3 and with previous 
studies at similar redshift. We also interpret our results 
in terms of AGN triggering mechanisms.

\subsection{Redshift evolution of the AGN bias}

Figure \ref{fig5} (right panel) shows the redshift evolution of the AGN bias
estimated using moderate luminosity AGN detected in different X-ray 
surveys. 
Interestingly, moderate luminosity AGN occupy DM halo masses
of log M$_h$ $\sim$ 12.5-13.5 M$_{\odot}$ h$^{-1}$ up to z$\sim$2,
%and therefore
%cluster like groups of galaxies.
%The general picture at z$<$2 is that the bias of moderate 
%luminosity X-ray AGN increases with redshift 
tracing a constant group-sized halo mass. 
Allevato et al. (2011) have shown that XMM COSMOS AGN 
(L$_{bol}$ $\sim$ 10$^{45.2}$ erg s$^{-1}$) reside in 
DM halos with constant mass equal to logM$_h$ = 13.12$\pm$0.07 
M$_{\odot}$ h$^{-1}$ up to z = 2. They also argue that this high bias can not 
be reproduced assuming that major merger between gas-rich 
galaxies (Shen 2009) is the main triggering scenario for moderate luminosity 
AGN. By contrast, at z$\sim$3, Allevato et al. (2014) 
found a drop in the mass of the hosting halos, with 
\textit{Chandra} and XMM-Newton COSMOS AGN
(L$_{bol}$ $\sim$ 10$^{45.3}$ erg s$^{-1}$), 
inhabiting halos of logM$_h$ = 12.37$\pm$0.10 M$_{\odot}$ h$^{-1}$.

In the present paper we measure a bias for 221.6 CCL AGN at 2.9$\leq$z$\leq$5.5
equal to b = 6.6$^{+0.6}_{-0.55}$, that corresponds to a typical mass of the hosting dark matter halos of logM$_h$ = 12.83$^{+0.12}_{-0.11}$ h$^{-1}$ M$_{\odot}$.
This result suggests a higher bias for CCL AGN compared to previous studies in COSMOS at z$\sim$ 3. In fact, Allevato et al. (2014) found a bias of 3.85$^{+0.22}_{-0.21}$ at $\langle z \rangle$=2.8, using a sample of XMM and \textit{Chandra} AGN.
Although the two samples only partially overlap, we argue that the most likely explanation of these differences lies in the small value  of $\pi_{max}$ (= 40 h$^{-1}$Mpc) used in Allevato et al. (2014). As shown in Figure 3, the bias strongly increases with $\pi_{max}$
due to the large phot-z errors and the use of 40 h$^{-1}$Mpc might produce an underestimated clustering signal.

To verify this effect, we took the same sample used in 
Allevato et al. (2014), i.e. 346 XMM and \textit{Chandra} COSMOS AGN with known spec or phot-z $>$ 2.2.
As already mentioned, Allevato et al. (2014) used the classic LS estimator where the phot-zs Pdfs are not taken into account. 
Using their same classic approach, we estimated the projected 2pcf for different values of $\pi_{max}$ and found that the clustering signal converges \textit{only} at $\pi_{max} \geq$ 200 h$^{-1}$Mpc. In particular, we derived a bias b = 5.8$^{+0.61}_{-0.65}$ for $\pi_{max}$ = 200 h$^{-1}$Mpc, which 
corresponds to a typical hosting halo mass logM$_h$ = 12.92$^{+0.13}_{-0.18}$.
As shown in Figure 5 (left panel) the bias estimated using $\pi_{max}$ = 200 h$^{-1}$Mpc is in agreement with the bias of CCL AGN at z$\sim$3.4 as derived in the present work. 
% and suggests, instead, a constant redshift 
%evolution of the bias factor at z$>$2.}
%Our results using \textit{Chandra} COSMOS-Legacy AGN
%at $\langle z \rangle$=3.4 suggest, instead,
%a larger bias, with X-ray AGN residing in halos with mass of the order of $\sim$ 6$\times$10$^{12}$ M$_{\odot}$ h$^{-1}$.
%As shown in Section 4, once a proper value of $\pi_{max}$ (= 200 h$^{-1}$Mpc) is used in Allevato et al. (2014), the %results are consistent and only a slight increase of the AGN bias can be observed between z$\sim$3 and $\sim$3.4. 
It is worth noticing that also the mean luminosity of the samples is increasing with z, 
with mean L$_{bol}=10^{45.5}$ erg/s for XMM and \textit{Chandra} COSMOS AGN at z$\sim$3 and $\sim$46 erg/s for CCL AGN at higher z. 

These results imply that at z$>$3: $i$) the typical hosting halo mass 
of moderate luminosity AGN remains almost constant with 
redshift, going from $\sim$8.3 $\times$ 10$^{12}$ at z=2.8 to $\sim 6.7 \times$ 10$^{12}$ M$_{\odot}$ h$^{-1}$ 
at z$\sim$3.4, since a lower mass is required to yield the same bias at a higher redshift; $ii$) 
moderate luminosity AGN still inhabit group-sized halos at high redshift, but slightly less massive than observed in different independent studies using X-ray selected AGN at z$\leq$2.

\begin{figure}
\plotone{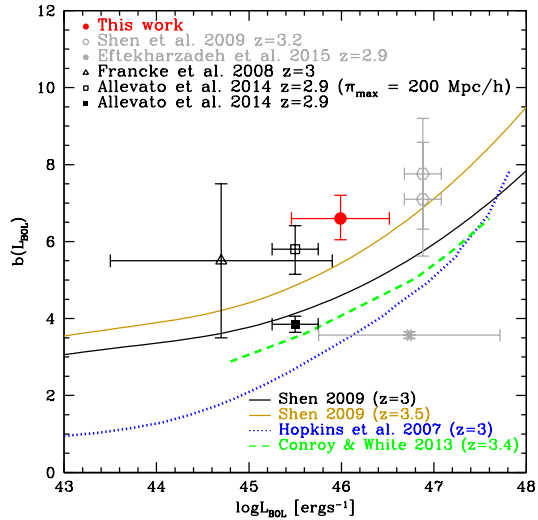}
\caption{\footnotesize Predicted bias as a function of bolometric luminosity, computed according to Shen (2009) at z = 3 (in black) and z = 3.5 (in gold), Hopkins et al. (2007b) at z=3 (blue dotted line) and Conroy \& White (2013) at z=3.4 (green dashed line) compared to previously estimated bias factors for optically selected quasars at z=3.2 (Shen et al. 2009,  gray hexagon), X-ray selected AGN at z=3 (Francke et al. 2008, open triangle), \textit{Chandra} and XMM-COSMOS AGN at z=2.9 as estimated in Allevato et al. (2014, filled black square) and as re-estimated in the present work (black empty square) and our results (red circle). The errors on the L$_{bol}$ axis correspond to the dispersion of the bolometric luminosity distributions for the different subsets. }
%The lines show the theoretical model which assumes a quasar phase triggered by a major merger as described in Shen 2009 (see the text for more details). \textbf{Bias factors from different studies are converted to a common cosmology ($\Omega_{\Lambda}$ = 0.7, $\Omega_m$ = 0.3, $\sigma$8 = 0.8).}} 
\label{fig6}
\end{figure}

\subsection{Previous studies at high redshift}

The evolution of the bias with redshit has been studied 
in Eftekharzadeh et al. (2015) for SDSS-III/BOSS quasars at 
2.2 $\leq$z $\leq$3.4. They investigated 
the redshift dependence of quasar clustering in 
three redshift bins and found no evolution 
of the correlation length and bias. In terms of halo mass, 
this corresponds to a characteristic halo mass that decreases with redshift, with halo masses 
of 3$\times$ (6$\times$) and $\sim$0.6$\times$ (1.3$\times$) 10$^{12}$ M$_{\odot}$ h$^{-1}$ at z$\sim$2.3 and $\sim 3$, 
respectively, where the dark matter halo masses are estimated using Tinker et al. 2010 (Sheth et al. 2001).

These results are surprisingly different in terms of bias and halo mass when compared to Shen et al.
(2009). At z$\sim$3, Eftekharzadeh et al. (2015) derived halo masses 
that are close to an order-of-magnitude smaller than those presented in 
Shen et al. (2009). In this latter study, they measured the bias of 
SDSS-DR5 quasars with mean L$_{bol}$ $\sim$ 10$^{47}$ erg/s, at 
$\langle z \rangle$ = 3.2.
Even if with large uncertainties, their results suggest 
that luminous quasars reside in massive halos with mass few times 
$10^{13}$ h$^{-1}$M$_{\odot}$ (based on Sheth et al. 2001). 
Although the two samples do not completely overlap, 
Eftekharzadeh et al. (2015) argue that the most 
likely explanation of these differences lies in the improvements in SDSS photometric 
calibration and target selection algorithms as well as in the much larger number of quasars that 
afford greater measurements precision compared to Shen et al. (2009).

Our results are in disagreement with the bias factor equal to 
b = 3.57$\pm0.09$ at z$\sim$3 derived in Eftekharzadeh et al. (2015). This disagreement might be due to the slightly different average redshift and the significantly different luminosity (almost one order of magnitude) 
of the samples used in the two different studies. An additional important difference between the samples is that 
our catalog of CCL AGN includes both Type 1 and Type 2 AGN, 
while Eftekharzadeh et al. (2015) use Type 1 BOSS luminous quasars. 
It is also worth noticing that in Eftekharzadeh et al. (2015) the bias is derived 
modelling the z-space 2pcf in the Kaiser formalism, i.e. not including the effect of 
random peculiar velocities and redshift errors.
The r$_0$ value (the correlation length of the projected 2pcf) reported in their Table 5 
would suggest, instead, a higher bias ($\sim$5) when derived assuming that w$_p$($r_p$) is modelled by 
a power law with index $\gamma$= -2 in the range r$_p$=4-30 h$^{-1}$ Mpc.

At a slightly lower redshift, Francke et al. (2008)
estimated the correlation function of a small sample of 
X-ray AGN with L$_{bol} \sim 10^{44.8}$ erg s$^{-1}$, 
in the Extended Chandra Deep Field South (ECDFS).
Given the small number of sources they only infer a minimum mass 
of halos hosting X-ray ECDFS AGN 
of logM$_{min}$ = 12.6$^{+0.5}_{-0.8}$ h$^{-1}$ M$_{\odot}$ (based on Sheth et al. 2001 formalism).
Our result is in agreement with this study at z$\sim$3,
but measured with higher accuracy.

Recently, Ikeda et al. (2015) investigated the clustering properties 
of low-luminosity quasars at z$\sim$4 using the cross-correlation function of 
quasars and LBGs in the COSMOS field. They estimated the bias 
factor for a spectroscopic sample of 16 quasars and a total sample of 25 quasars including sources with photometric 
redshifts. They obtained a 86\% upper limit for the bias of 5.63 and 10.50 for the total and 
spectroscopic sample, respectively.

%that moderate
%luminosity X-ray AGN reside in dense environment (logM$_h \sim$ 13  
%M$_{\odot}$ h$^{-1}$) up to z$\sim$3.4,
%as already found for XMM-COSMOS AGN at z$<$2 (Allevato et al. 2011).
%This indicates that merger models fail in reproducing the 
%clustering properties of X-ray AGN not only at z$<$2 but up 
%to z$\sim$ 3.4, and that the 
%nature of X-ray AGN triggering is not changing at high redshift.

%there is mounting observational evidence suggesting that moderate levels of AGN activity might not be always casually connected to galaxy interactions. Theoretically, in-situ processes, such as disk instabilities or stochastic accretion of gas clouds, have also been invoked as triggers of AGN activity 

\subsection{Comparison to theoretical models}

Figure 6 shows the predicted evolution of the 
AGN bias as a function of the bolometric luminosity, computed 
according to the framework of the growth and evolution of BHs 
presented in Shen (2009, see also Shankar 2010) 
at z = 3 and 3.5 Their model assumes that quasar activity 
is triggered by major mergers of host halos 
(e.g. Kauffmann \& Haehnelt 2000).

The major merger model is quite successful in predicting the bias of 
COSMOS AGN at z=2.8 as presented in Allevato et al. (2014), but underpredicts 
the bias re-estimated in the present work using the same AGN sample and $\pi_{max}$ = 200 h$^{-1}$ Mpc.
Given the large error bars, the model is in broad agreement
with the bias of luminous quasars at similar redshift as measured in Shen et al. (2009)
and X-ray AGN as estimated in Francke et al. (2008). 

The prediction from the model slightly underpredicts 
our results for CCL AGN at z$\sim$3.4.
We verified that the mismatch between merger models and our data does not change
if a few parameters, such as the light curve
or the host halo mass distribution are changed 
in the major merger model. In fact, our result is still not well reproduced 
%The dotted lines in Figures 5 marks
by the predictions from a modified Shen (2009) model in
which the post-peak descending phase is cut out, with
all other parameters held fixed. 
%In addition, the continuous line shows a variant 
On the other hand, a model characterized by a steepening
in the L$_{peak}$-M$_h$ relation 
%below M$_h$ $\simeq$10$^{12}$ M$_{\odot}$/h,
%with L$_{peak}$ $\propto$ M$_h^{5}$, that 
mainly implying that preferentially 
lower-luminosity quasars are now mapped to more massive, less
numerous host DM halos, still underpredicts our results.
%Interestingly, the major merger model overpredics the bias of 
%very luminous SDSS-III/BOSS quasars, that has been estimated with very 
%high accuracy in Eftekharzadeh et al. (2015).}

A similar mismatch has also been found for a sample of CCL AGN at z=3-6.5 
in terms of observed number counts (Marchesi et al. 2016b). In fact, they 
verified that the reference model overproduces the observed number counts 
by a factor of 3 to 10, depending on the redshift.
%The triggering mechanisms of AGN are still debated. Some of the most 
%popular ones include galaxy merging. 

We also compare the observations with the theoretical model presented in 
Hopkins et al. (2007b), that adopts the feedback-regulated quasar light-curve/lifetime models 
from Hopkins at al. (2006) derived from numerical simulations of galaxy mergers that 
incorporate BH growth.
Even if we assume an evolution with redshift, this model underpredicts 
the bias factor of CCL AGN. 

A similar tension is also observed 
when comparing with the semi-empirical model presented in Conroy \& White (2013).
In the latter, the BH mass is linearly related to galaxy mass and 
connected to dark matter halos via empirical constrained relations.
This model makes no assumption about what triggers the AGN activity 
and includes a scatter in the AGN luminosity - halo mass relation, 
contrary to Hopkins et al. (2007b) and Shen (2009).
Conroy \& White (2013) show that this semi-empirical model naturally 
reproduces the clustering properties of quasars at z$<$3, but shows some 
tension at higher redshift. They argue that this disagreement can be explained if AGN have a 
duty cycle close to unity at z$>$3, indicating that we approach the era of 
rapid BH growth in the early universe.

Recently, Gatti et al. (2016) 
have used advanced semi analytic models (SAMs) of galaxy 
formation, coupled to halo occupation modelling, to investigate AGN triggering 
mechanisms such as galaxy interactions and disk instabilities.
They compared the predictions with high
redshift clustering measurements from Allevato et al. (2014), Shen et al. (2009) 
and Eftekharzadeh et al. (2015). 
Their SAMs underpredict the bias of luminous quasars shown in
Shen et al. (2009). The mismatch is reduced when the models are compared to 
Eftekharzadeh et al. (2015). 
They pointed out that, 
irrespective of the exact implementations in their SAMs, at low-z moderate-luminosity AGN 
(L$_{bol}\sim$10$^{44-46}$ erg/s) mainly inhabit halos 
with mass $\sim$10$^{12-13}$ M$_{\odot}$ for both galaxy interaction and 
disk instabilities models (even if disk instabilities do not trigger the most luminous AGN with 
L$_{bol}\geq$10$^{47}$ erg/s). At higher redshift (z$\sim$2.5), structures with mass greater than $M_h>10^{13}$
M$_{\odot}$ become significantly rarer, relegating active galaxies to live mainly in less massive environment.
Moreover, in all models only galaxies with stellar masses above 10$^{11}$ M$_{\odot}$ 
would be able to host AGN with luminosity of L$_{bol}\sim$10$^{46}$ erg/s and highly biased 
such as COSMOS AGN at z$>$2-3. This would imply that the 
characteristic M$_{star}/M_{h}$ ratio in AGN hosts should 
increase with lookback time, as expected from basic considerations 
on number densities evolution between the halo mass function and AGN luminosity function 
(e.g., Shankar et al. 2010).

\section{Conclusions}
\label{sec:conc} 

We use the new CCL catalog to probe the projected and redshift-space 2pcf 
of X-ray selected AGN for the first time at 2.9$\leq$z$\leq$5.5, using the generalized 
clustering estimator based on phot-z Pdfs in addition to any
available spec-z.
%with high accuracy at scales r$_p$ = 0.1-30 Mpc h$^{-1}$.
%We estimate the 2pcf of 221.6 Chandra COSMOS-Legacy AGN at 2.9$<$z$<$5.5, 
We model the clustering signal with the 2-halo model 
and we derive the bias factor and the typical mass 
of the hosting halos.
Our key results are:

\begin{enumerate}

\item At z$\sim$3.4, CCL 
AGN have a bias b = 6.6$^{+0.60}_{-0.55}$, which 
corresponds to a typical mass of the hosting halos 
of log M$_h$ = 12.83$^{+0.12}_{-0.11}$ 
h$^{-1}$ M$_{\odot}$. A similar bias 
%b = 6.53$^{+0.52}_{-0.55}$ 
is derived using the z-space 2pcf, modelled including the typical phot-z error 
$\sigma_z$ = 0.052 of our sample. This confirms that the convergence of
the projected 2pcf observed only at large scales ($\pi_{max}\geq200$ h$^{-1}$ Mpc) 
is due to large phot-z errors.

\item A slightly larger bias b = 7.5$^{+1.6}_{-1.7}$ 
(but consistent within the error bars) is found using a 
sample of 107 CCL AGN with known spec-z. 
The modelling of $\xi(s)$ suggests that this larger bias can 
be explained assuming that spec-zs are affected by errors of 
the order of $\sigma_z = 0.02-0.025$. This would explain the convergence of the 
projected 2pcf surprisingly observed only at $\pi_{max}\geq200$ h$^{-1}$ Mpc, 
even when phot-zs are not included in the analysis. However, 
given the low statistics smaller spec-z errors and then bias 
can not be ruled-out.

\item We estimate the bias factor for the sample of 346 XMM and \textit{Chandra} AGN used in 
Allevato et al. (2014) using $\pi_{max}=200$ h$^{-1}$ Mpc in estimating the projected 
2pcf and then accounting for the large phot-z errors. In particular we found b = 5.8$^{+0.61}_{-0.55}$, which 
is significantly larger than the AGN bias measured in Allevato et al. (2014) and corresponds to 
logM$_h$ = 12.92$^{+0.13}_{-0.18}$ at z=2.8.

\item Our results suggest only a slight increase of the bias factor of COSMOS AGN at z$\gtrsim$3,
with the typical hosting halo mass of moderate luminosity
AGN almost constant with redshift and equal to 
logM$_h$ = 12.92$^{+0.13}_{-0.18}$ at z=2.8 and log M$_h$ = 12.83$^{+0.12}_{-0.11}$ 
at z$\sim$3.4, respectively.
 
\item The observed redshift evolution of the bias of COSMOS 
AGN implies that moderate luminosity AGN still inhabit group-sized halos,
%and that a redshift evolution tracing a constant halo mass track is almost still valid at z$\gtrsim$3,
but slightly less massive than observed in different independent studies using X-ray AGN at z$\leq2$.

\item Theoretical models presented in Shen (2009) and Hopkins et al. (2007b) that assume an AGN activity
mainly triggered by major mergers of host halos underpredict our results at z$\sim$3.4 for 
CCL AGN with mean L$_{bol} \sim 10^{46}$ erg s$^{-1}$.
A similar tension is also observed when comparing to the semi-empirical models presented in Conroy \& White (2013).
In the latter model, this disagreement can be explained if AGN have a duty cycle approaching unity at z$>$3.
On the other hand, following the semi-analytic models presented in Gatti et al. (2016),
in both galaxy interaction and disk instability models
%moderate-luminosity AGN (L$_{bol} \sim 10^{44-46}$ erg/s) at z$\sim$2-3, 
only galaxies with stellar masses above 10$^{11}$ M$_{\odot}$ 
would be able to host AGN with luminosity of L$_{bol}\sim$10$^{46}$ erg/s and highly biased 
such as COSMOS AGN at z$>$2-3.
\end{enumerate}

Only future facilities, like the X-ray Surveyor (Vikhlinin 2015) and Athena (PI K. P. Nandra), 
will be able to collect sizable samples ($\sim$1000s) of low luminosity (L$_X<$10$^{43}$ erg/s) 
AGN at z$>$3 (Civano 2015), allowing to explore the clustering for significantly less luminous source 
and to test AGN triggering scenarios at different AGN luminosities.

\acknowledgments

We thank the anonymous referee for helpful comments.
We gratefully thank Federico Marulli for helpful discussions and inputs.
We acknowledge the contributions of the entire COSMOS 
collaboration consisting of more than 100 scientists. More information 
on the COSMOS survey is available at http://www.astro.caltech.edu/COSMOS. 
VA and AF wish to acknowledge Finnish Academy award, decision 266918.
This work was supported in part by NASA Chandra grant number GO3-14150C and also GO3-14150B  (F.C., S.M., V.A., H.S.). 
RG acknowledge receipt of NASA Grant NNX15AE61G.
TM is  supported by UNAM-DGAPA Grant PAPIIT IN104216 and CONACyT Grant Cientifica Basica 179662.
KS gratefully acknowledges support from Swiss National Science Foundation Grant PP00P2\_138979/1.

\clearpage

\end{document}